# Designer magnetic topological graphene nanoribbons

*Shaotang Song[1,7], Pei Wen Ng[1,7], Shayan Edalatmanesh[2,6,7], Andrés Pinar Solé[2], Xinnan Peng[1], Jindřich Kolorenč[2], Zdenka Sosnová[2], Oleksander Stetsovych[2], Jie Su[1], Jing Li[1], Hongli Sun[3], Alexander Liebig[4], Chenliang Su[3], Jishan Wu[1], Franz J. Giessibl[4], Pavel Jelinek[2,6*], Chunyan Chi[1*], Jiong Lu[1,5*].*

[1]Department of Chemistry, National University of Singapore, 3 Science Drive 3, Singapore 117543, Singapore.

[2]Institute of Physics of the Czech Academy of Sciences, Prague, Czech Republic.

[3]SZU-NUS Collaborative Center and International Collaborative Laboratory of 2D Materials for Optoelectronic Science & Technology of Ministry of Education, Engineering Technology Research Center for 2D Materials Information Functional Devices and Systems of Guangdong Province, Institute of Microscale Optoelectronics, Shenzhen University, Shenzhen 518060, People's Republic of China.

[4]Department of Physics, University of Regensburg, Regensburg 93053, Germany.

[5]Institute for Functional Intelligent Materials, National University of Singapore, 117544, Singapore.

[6]Regional Centre of Advanced Technologies and Materials, Czech Advanced Technology and Research Institute (CATRIN), Palacký University Olomouc, 78371 Olomouc, Czech Republic.

[7]These authors contributed equally: Shaotang Song, Pei Wen Ng, Shayan Edalamanesh.

*e-mail: jelinekp@fzu.cz; chmcc@nus.edu.sg; chmluj@nus.edu.sg

**Abstract**

The interplay of magnetism and topology lies at the heart of condensed matter physics, which offers great opportunities to design intrinsic magnetic topological materials hosting a variety of exotic topological quantum states including the quantum anomalous Hall effect (QAHE), [1] axion insulator state, [2,3] and Majorana bound states. [4-7] Extending this concept to one-dimension (1D) systems offers additional rich quantum spin physics with great promise for molecular-scale spintronics. [8] Despite recent progress in the discovery of symmetry-protected topological quantum phases in 1D graphene nanoribbons (GNRs), [9-13] the rational design and realization of magnetic topological GNRs (MT-GNRs) represents a grand challenge, as one must tackle multiple dimensions of complexity including time-reversal symmetry (TRS), spatial symmetry (width, edge, end geometry) and many-electron correlations. [14] Here, we devised a new route involving the real- and reciprocal-space descriptions by unifying the chemists' and physicists' perspectives, for the design of such MT-GNRs with non-trivial electronic topology and robust magnetic terminal. Classic Clar's rule offers a conceptually qualitative real-space picture to predict the transition from closed-shell to open-shell with terminal magnetism, and band gap reopening with possible non-trivial electronic topology in a series of wave-like GNRs, which are further verified by first principle calculations of band-structure topology in a momentum-space. With the advance of on-surface synthesis and careful design of molecular precursors, we have fabricated these MT-GNRs with observation of topological edge bands, whose terminal π-magnetism can be directly captured using a single-nickelocene spin sensor. [15-17] Moreover, the transition from strong anti-ferromagnetic to weak coupling (paramagnetism-like) between terminal spins can be controlled by tuning the length of MT-GNRs. Our joint chemistry and physical approach enables the design and realization of MT-GNRs with non-trivial topology and robust magnetism, which represents a new class of quantum matter, crucial for exploration of 1D quantum spin physics and quantum information technology.

**Main**

The integration of topological quantum phases and intrinsic magnetism in three-dimensional (3D) and two-dimensional (2D) topological insulators (TIs) lies in the interplay between topology and symmetry.[1-7] Topological quantum phases are characterized by the topological invariant ($\mathbb{Z}_2$) corresponding to integration of the Berry curvature of the occupied energy bands for each spin species in the Brillouin zone (BZ), and protected by time-reversal symmetry (TRS), which can be broken in the presence of magnetism.[1,18,19] A broken TRS by introducing intrinsic magnetism from magnetic layers and magnetic dopants leads to the formation of an exchange gap in topologically-protected otherwise gapless Dirac surface states, allowing the realization of the QAHE and axion insulator state in 3D and 2D TIs. [20,21]

The concept of topological quantum phases has long been reported in 1D polyacetylene described by Su-Schrieffer-Heeger (SSH) model, and recently been discovered in 1D acene-motif-derived polymers *via* tailoring their π-conjugation, as well as 1D GNRs through on-surface synthesis [9-13,22-28]. The topological invariant of these 1D polymers and GNRs ($\mathbb{Z}_2$) is characterized by the sum of the intercell Zak phase of the occupied bands, an integral of the Berry connection in the 1D BZ (either $\mathbb{Z}_2$=0 or $\mathbb{Z}_2$=1, corresponding to trivial and non-trivial topology).[29,30] Incorporation of magnetism into 1D topologically non-trivial GNRs may allow to design the spin-selective transport channel, symmetry-protected spin-correlated end state or Majorana bound states in close proximity to a superconductor, crucial for developing molecular-scale spintronics. In contrast to 2D or 3D counterparts, the topological phase in 1D GNRs is also dictated by their spatial symmetries defined by width, edge, and end geometry, in addition to TRS.[9] Moreover, 1D topological GNRs are also predicted to support half-filled topological end states localised at the GNRs-vacuum boundary characterized by different values of $\mathbb{Z}_2$. It has been predicted that inclusion of strong on-site Coulomb repulsion can split such localized topological end states into spin-up and spin-down polarized states, resulting in the emergence of π-magnetism.[23,25,31] However, these many-electron correlated effects are often beyond the theoretical framework description from tight-binding (TB) models and Density Functional Theory (DFT) within local density approximation. Therefore, rational design of MT-GNRs has been challenging as multiple dimensions of complexity including TRS, spatial symmetry and strong electronic correlation need to be tackled.

**The design concept of MT-GNRs**

To circumvent these challenges, we devised a new strategy for the rational design of MT-GNRs with non-trivial electronic topology and magnetic terminal from both chemists' and physicists' perspectives (Fig. 1a). Since magnetic and topological quantum phases in GNRs are dictated by π electrons, we first adopted the classic Clar's rule based on "local" viewpoint of π sextets in aromatic resonance structure as a conceptually qualitative real-space picture to predict the presence of band gap reopening in a series of wave-like GNRs through trans-coupling of zigzag-edged nanographene unit [denoted as (*a,b*)-tzGNRs), *a* and *b* represent the number of the two edges' zigzag lobes of the parallelogram nanographene as fundamental building block.] (Fig. 1a). According to Clar's rule, a resonance structure of polycyclic aromatic hydrocarbons or graphene nanostructures with the maximal number of nonadjacent π sextets represents the most stable form or the major resonance contributor.[32,33] Increasing the ratio of nonadjacent π sextets gains more aromatic resonance energy, which favors the formation of a larger energy gap in this series of tzGNRs. The ratio of Clar π sextets first decreases from 33.33% for (1,3)-tzGNR to 28.57% for (2,3)-tzGNR, and then increases to 31.82%, 33.33%, and 31.58% for (3,3)-, (4,3)- and (5,3)- tzGNRs, respectively, suggesting a band gap reopening as GNRs transits from (2,3) to (3,2), with possible emergence of non-trivial electronic topology in this series of tzGNRs. Moreover, such a band gap reopening is also accompanied by the variation of the maximal number of Clar π sextets in unit cell of the tzGNRs between the closed-shell and open-shell resonance structures (Fig. 1a). As such, the Clar's rule predicts a transition from closed-shell to open-shell character as the width of tzGNR increases, which favors the formation of unpaired electrons at terminal of tzGNRs (a ≥ 3) as the maximal number of π sextets are significantly higher in open-shell than closed-shell structures. [34]

Therefore, Clar's rule offers a facile and qualitative evaluation (chemist's perspective) to predict such exotic quantum states in GNRs, which can be further verified from physicists' perspective based on band-topology calculations of these GNRs as a function of their width and lengths. DFT calculations also reveal that the band gap of infinite tzGNRs first decreases from 2.86 eV for (1,3) to 0.88 eV for (2,3) and then reopens from 1.75 eV for (3,3) to 1.3 eV and 0.48 eV for (4,3) and (5,3) respectively (Fig. 1c and 1e). Such a reopening is also accompanied by level crossing of the conduction band (CB) and valence band (VB) wavefunctions at the Γ-point (see Extended Data Fig. 4). This level crossing causes the topological phase transition from topologically trivial to non-trivial phases. Moreover, the transition is also confirmed by calculations of Zak-phase using DFT band structure of GNRs (Fig. 1e) indicating the change of $\mathbb{Z}_2$ from 0 (m≤2) to 1 (a≥ 3). [12,26]

**On-surface synthesis and characterization of the tzGNRs**

Anthracene was selected as the building block for the synthesis of (1,3)-tzGNR (Extended Data Fig. 1). In addition, two dimethylphenyl groups were installed onto anthracene and anthanthrene to target the wider (3,3)-tzGNR and (4,3)-tzGNR. For all the designed monomers, two Br atoms were then attached onto their peripheries as active sites for polymerization. The detailed synthetic route is depicted in supplementary information (Scheme S1). The on-surface synthesis first involves the sublimation of precursors onto a clean Au(111) surface, subsequent stepwise thermal precursor activation (dehalogenation), polymerization, and finally cyclodehydrogenation of the polymer at given temperatures (Extended Data Fig. 1). Despite limited control over the selectivity between homochiral and heterochiral polymerization, the targeted GNRs in trans-coupling can be obtained with a moderate yield (Extended Data Fig. 2-3).

We first used the bond-resolved scanning tunnelling microscopy (BR-STM) technique with a CO functionalized tip to image the atomic structure of (a, 3)-tzGNRs at the constant-height mode (Fig. 2a-c). Local electronic structures of the (1,3)-, (3,3)- and (4,3)-tzGNR were then characterized using d$I$/d$V$ point spectroscopy (Fig. 2). Spectra collected at the armchair (blue) and zigzag (red) segment regions of tzGNRs reveal both prominent empty and occupied electronic states at VB and CB sides, which are labelled as A and B, respectively. The band gaps of the (1,3)-, (3,3)-, and (4,3)-tzGNR are determined to be 2.45 eV, 2.20 eV, and 1.25 eV, respectively, consistent with the trend predicted by DFT calculations (Fig. 1c and 1e). d$I$/d$V$ maps recorded at energetic position of peak B (CB side) of all the three GNRs reveal the same characteristic bright lobes feature localized at the alternating armchair regions. In contrast, the d$I$/d$V$ maps acquired at energetic position of peak A (VB side) show different features at these regions. In addition, there are two states localized at the middle of the anthracene zigzag lobes of the (1,3)-tzGNR, which are absent for the (3,3)- and (4,3)-tzGNRs (Fig. 2d-f, and Fig. S1-S3). The calculated d$I$/d$V$ maps of all three GNRs at the energetic positions near band edge of both CB and VB sides (see Figs. S1-3), match well with experimental ones, confirming the validity of DFT method to properly describe the electronic structure of GNRs and their corresponding topological phase.

We then carried out a detailed characterisation of the terminal of (3,3)- and (4,3)-tzGNRs to probe their end geometry and magnetism (Fig. 3 and Extended Data Fig. 3). In contrast to the honeycomb pattern observed in the center of the (4,3)-tzGNRs, terminals show an enhanced contrast with lobe features in the constant-current BR-STM image recorded at 2 mV (Fig. 1d), suggesting a LDOS enhancement in the vicinity of Fermi level ($E_F$). To resolve the atomic structure of the terminal units, we then performed the constant-height frequency-modulation ncAFM measurement using the qPlus sensor with a CO-functionalized tip (refer to methods section).[35,36] The ncAFM image of the end of (4,3)-tzGNR clearly resolves its intact honeycomb pattern (Fig. 3a). On the other hand, the BR-STM images of the terminal of (3,3)- and (4,3)-tzGNR acquired at 20 mV at constant height mode (Extended Data Fig. 3c, Fig. 4d), show tail features around the periphery of the honeycomb lattices, consistent with the spin-density distribution calculation (Fig 4e), which indicates delocalized spin populations near the terminal [37].

The spectra acquired over the bright sites of the terminal monomer units in both (3,3)- and (4,3)-tzGNRs show pronounced zero-bias peaks (ZBPs) (Extended Data Fig. 3, Fig. 3d, and Fig. 4c). Such ZBP can be attributed to a Kondo resonance due to the screening of the local magnetic moment, as it exhibits a characteristic temperature-dependent broadening wherein the half-width at half-maximum value can be fitted to the Fermi-liquid model with a Kondo temperature of $T_k = 35 \pm 0.2$ K (Fig. 3d,e).[38] The intensity of the Kondo resonance dominates over the first few lattice sites within the last monomer unit but fades rapidly as moving towards the interior GNR bulk, consistent with experimental and simulated d$I$/d$V$ mapping at 0 mV of (4,3)-tzGNR (Fig. 3b, 3c). This result indicates that the spin is delocalized over the last unit of the terminal, highlighting sufficient stability to prevent spin quenching upon absorption of the metal surface. [34]

To further probe the magnetic state at the terminal unit, we employed inelastic electron tunneling spectroscopy (IETS) using nickelocene (NiCp$_2$) functionalized tips.[15-17] Namely, we used the characteristic d$^2I$/d$V^2$ signal corresponding to the spin-flip of NiCp$_2$ spin ($S_{Ni}$ = 1) to sense a local magnetic moment at the terminal of the GNRs (Fig. 3f). The d$^2I$/d$V^2$ signal of NiCp$_2$-tip consists of a peak/dip symmetrically located around zero bias (Fig. 3g). These features are attributed to the spin-flip excitation of NiCp$_2$ from the ground state ($m_s$ = 0) to the double degenerate excited states ($m_s$ = ± 1).[17] We first acquired the d$^2I$/d$V^2$ spectra over interior bulk regions of (4,3)-tzGNRs at different tip-sample distances, revealing a characteristic peak (dip) at + 5 meV (- 5 meV) bias (Fig. 3g). At a reduced tip-sample distance (from 0 to 50 pm, with an incremental step of 10 pm), the intensity of the peak (dip) increases but the energetic position and the width of the peak/dip remains constant, suggesting an absence of magnetic interaction between the bulk GNR region and the NiCp$_2$-tip. The presence of an exchange-field associated with the local magnetic moment ($S_{end}$ = 1/2) at the GNR terminal may modulate the IETS signal of the NiCp$_2$-tip. To confirm this hypothesis, we acquired a series of d$^2I$/d$V^2$ spectra above the terminal GNR unit (Fig. 3h) in the same tip-sample distance range as over the bulk region (Fig. 3g). Indeed, the set of d$^2I$/d$V^2$ spectra taken over the terminal of (4,3)-GNR show a monotonic shift of the peak (dip) accompanied by its broadening when reducing the tip-sample distances.

To gain more insight into the origin of the observed behaviour, we carried out IETS simulations.[31] To model the interaction between the local magnetic moment of NiCp$_2$-tip ($S_{Ni}$ = 1) and the GNR end ($S_{end}$ = 1/2), we solved a two-site Hubbard model consisting of the partially-filled 3$d^8$ shell of the NiCp$_2$ and one filled orbital with a single electron (S = 1/2) representing for the end state, respectively. The variation of the tip-sample distance is represented by the modulation of an effective hopping *t* between the two sites (refer to [31] for more details about the IETS model). The IETS signal of simulated spectra remains unaltered in far tip-sample distances (Fig. 3k). However, two degenerate excited states ($m_s = \pm 1$) gradually split (Fig. 3j and 3k) and vary in intensity as decreasing the tip-sample distance (corresponding to a larger effective hopping *t*). This splitting is caused by the magnetic interaction of NiCp$_2$-tip with the local magnetic moment $S_{end} = 1/2$. As the tip approaches further towards the sample, the energy splitting becomes larger and the intensity of the new peak/dip is enhanced (Fig. 3k). Note that the energy splitting is much smaller than the energy resolution of our experimental IETS, which prevents a direct visualization of both peaks. Instead, if a finite energy width of each peak (1.7 meV FWHM) is adopted, the simulated IETS spectrum resembles the experimental observation well, i.e. broadening and gradually shifting of the peak towards higher energies.

To better understand the origin of the evolution of the IETS signal upon the NiCp$_2$-tip approach, we analysed the magnetic interaction employing the spin Heisenberg model in the form:

$$\hat{H} = D_{Ni}(\hat{S}_z)^2 + J\hat{S}_{Ni}\hat{S}_{end} \qquad (1)$$

which accurately approximates the two-site Hubbard model as long as charge transfer between the sites is negligible. Parameters $D_{Ni}$ represents magnetic anisotropy of NiCp$_2$ and *J* is magnetic exchange coupling between the spin moment on NiCp$_2$ ($\hat{S}_{Ni}$) and the local spin at the terminal ($\hat{S}_{end}$). Fig. 3i compares solutions of the Hubbard and Heisenberg models, respectively. The perfect match between the models using the value of $J = 0.197t^2$ justifies the validity of the Heisenberg model (Fig. 3i and 3j). The Heisenberg model can be solved exactly, and the resulting three eigenstates $\Psi_i$ (i = 0-2) (see Extended Data Table 1) can be directly related to the individual peaks observed in the IETS signals. Consequently, they provide the physical origin of the individual peaks at different tip-sample distances. According to the analysis, we found that the degenerated excited states ($m_{Ni} = \pm 1$) of NiCp$_2$-tip combine with the local spin moment S = 1/2 to form two doubly degenerate states, one with the total z-projection of $m_{\text{tot}} = \pm 1/2$ and the other one with $m_{\text{tot}} = \pm 3/2$. This split explains the emergence of two peaks in the IETS spectra at small tip-sample distances. This agreement between the experimental and theoretical IETS data convincingly proves the presence of the local magnetic moment at the terminal of MT-GNRs.

Apart from the width-dependent behaviour of tzGNRs, Clar's rule also can predict the transition from closed-shell to open-shell character as a function of length. Taking (4,3)-tzGNRs as an exemplary case, the closed-shell Kekulé structure of these GNRs with finite length host 3$n$ aromatic sextets as a function of length ($n$ denotes the number of monomer units in the GNRs). Maximizing the aromatic sextets to 5$n$ – 1 creates their open-shell biradical forms with unpaired electrons at the terminal of the GNRs (Extended Data Fig. 7). Therefore, the open-shell non-Kekulé structure of (4,3)-tzGNRs consists of an extra number (2$n$-1) of π sextet as compared to their closed-shell counterpart (Extended Data Fig. 7). The additional Clar sextet rings in the biradical form gains extra aromatic stabilization energy and thus favor the formation of open-shell singlet ground state, wherein two spins are coupled anti-ferromagnetically (eg. dimer). However, the monomer ($n$ = 1) carries only one more π sextet in the open-shell structure, as compared to that of the closed-shell Kekulé structure. As a result, the stabilization energy is not sufficient to drive the monomer towards the open-shell form and thus adapts a closed-shell ground state. Such a trend is also demonstrated in zigzag-edged nanographene, where the [4]rhombene shows the closed-shell ground state in contrast to the open-shell ground state of the larger [5]rhombene on Au(111). [39]

To verify these predictions, we use precursors **5** and **4**, to synthesize the monomer and dimer, respectively (Scheme1 in supplementary information). The combination of **4** and **3** with controlled ratio enables the fabrication of (4,3)-tzGNRs with different lengths (Fig. 4d, Extended Data Fig. 6-7), allowing to probe the transition from non-magnetic to magnetic, and magnetic coupling between two opposite terminals. As shown in Fig. 4c and Extended Data Fig. 5, the (4,3)-tzGNR monomer indeed shows closed-shell ground state as evidenced by featureless d$I$/d$V$ spectra near E$_F$. The even smaller (3,3)-tzGNR monomer derived from precursor **2** (Extended Data Fig. 5) also adopts the same closed-shell structure, as evidenced by rather featureless d$I$/d$V$ signal near E$_F$. [40]

We then performed d$I$/d$V$ and IETS in the vicinity of E$_F$ to probe the spin-interactions of the GNR dimer and oligomers. d$I$/d$V$ spectrum taken at the dimer end shows a gap-like feature cross E$_F$, and the corresponding IETS spectrum reveals asymmetric conductance peak and dip at 35 and -35 meV, respectively (Fig. 4c and Fig. S5). These features can be attributed to the inelastic spin-flip excitation from singlet ground state to triplet excited state, [39,41-44] suggesting that the two spins at the terminal of the dimer are antiferromagnetically coupled (AFM) with an exchange coupling strength ($J_{eff}$) of 35 mV. The AFM coupling between two ends becomes negligible (paramagnetic-like behaviours) as the length of GNRs reaches three-unit cells, wherein Kondo resonance peak can be observed in trimer (Fig. 4c and Extended Data Fig. 6b). Such a length dependent spin-interaction is consistent with the trend predicted by spin-polarised DFT calculations, namely, a monotonic decrease of AFM coupling strength between terminal spins as the length of GNR increases (Fig. 4b). We also note that as-predicted coupling strength is slightly larger than the experimental value, which can be attributed to the metallic substrate screening.[45-47]

Therefore, the coexistence of magnetism and topological quantum phases in MT-GNRs can be well described by combining Clar's rule and spin polarized DFT calculations. The bulk-boundary correspondence in topological GNRs results in a symmetry-protected-topological (SPT) half-filled metallic end state. Strong electronic correlation effects split such states into spin-up and spin-down end states close to the conduction and valence band edges, respectively (Fig. S4), giving rise to local magnetic moment and its associated zero-bias Kondo resonance peak. The magnetic state is further verified by measuring the spin-flip IETS signal of the magnetic sensor (NiCp2-tip) showing a striking variation in the IETS spectra at the terminal of GNR. Therefore, the magnetism and topological quantum phases are interlocked in designer MT-GNRs.

**Conclusion**

We have demonstrated a new approach by unifying the chemists' and physicists' perspectives, to provide a powerful and predictive guideline for the rational design of MT-GNRs with non-trivial topology and π-magnetism. The Clar's rule offers a facile and qualitative evaluation to search such exotic quantum states in GNRs, which are further verified in momentum-space band topology calculations. With the advances of on-surface bottom-up synthesis and careful design of molecular precursors, we have fabricated these MT-GNRs with direct observation of π-magnetism at the terminal of MT-GNRs using a single-molecule spin sensor. Moreover, the transition from strong anti-ferromagnetic to weak coupling between terminal spins can be controlled by tuning the length of MT-GNRs. Our designer approach enables the experimental realization, and manipulation of terminal π-magnetism in MT-GNRs with non-trivial topology, which may create a major paradigm shift towards spintronic applications in quantum technologies.

**Methods**

**Precursor synthesis and GNR growth.** Detailed synthesis and characterization of precursors **1-5** are described in the Supplementary Information. Au(111) single crystal (MaTeck GmbH) was cleaned by multiple cycles of $Ar^+$ sputtering and annealing. Knudsen cell (MBE-Komponenten GmbH) was used for the deposition of precursor molecules onto clean Au(111) surfaces under ultrahigh vacuum conditions (base pressure, <2 × $10^{-10}$ mbar) for on-surface synthesis of a series of zt-GNRs. After deposition of precursors, the sample was annealed at elevated temperatures as stated in the main text for 20 min to induce radical polymerization and subsequent cyclodehydrogenation for the synthesis of tzGNRs.

**STM and d$I$/d$V$ characterization.** The experiments were conducted in LT-STM/AFM systems operated under ultrahigh vacuum (base pressure, P < 2 × $10^{-11}$ mbar) at a temperature of T = 4.5 K (Scienta Omicron) and 2.8 K (JT-SPECS), respectively. The d$I$/d$V$ spectra were collected using a standard lock-in technique with a modulated frequency of 479 Hz. The modulation voltages for individual measurements are provided in the respective figure captions. The STM tip was calibrated spectroscopically against the surface state of Au(111) substrate. All the bond-resolved STM images were taken at constant height scanning mode ($V$ = 20 mV) with a CO functionalized tip. The CO molecule was picked up by a sharp metal tip following a routine procedure.[48] The STM images were analysed and processed with Gwyddion software.[49]

The ITES ($d^2I/dV^2$) spectra taken with a nickelocene-functionalized tip were conducted using a SPECS LT-SPM at 2.8 K. For the deposition of nickelocene molecules onto the surface, the crucible was pre-pumped for ten minutes. Nickelocene was then introduced at a pressure of 5 ×$10^{-6}$ mbar through a leak valve and dosed directly to the cold sample in the STM chamber for a few seconds. Nickelocene tip functionalization was performed by scanning an isolated Nickelocene molecule in the constant current STM mode (set-point of $V_S$: 1-2 mV, $I$: 10 pA) with a Kolibri sensor[50]. The quality of the Nickelocene tips is first tested on Au(111) surface and only stable Nickelocene tips are used for the spectroscopy measurements.

The microscope of Scienta Omicron was equipped with qPlus sensors S0.8 with a resonance frequency of $f_0$ = 39.646K Hz, a stiffness of 3600 N/m, and a quality factor of 77394. nc-AFM images were collected at a constant-height frequency modulation mode using an oscillation amplitude of A = 50 pm. The tip-sample distance with respect to an STM set point is indicated in the figure caption the AFM image.

**DFT calculations.** DFT calculations for all free-standing finite and infinite systems (N = 1, 2, 3, 4 GNRs) were carried out using the FHI-AIMS and Fireball DFT packages. All geometry optimizations and electronic structure analyses were performed using the hybrid exchange–correlation functional PBE0. Systems were allowed to relax until the remaining atomic forces reached below $10^{-2}$ eV Å$^{-1}$. For all infinite systems with a periodic boundary condition, a Monkhorst–Pack grid of 18×1×1 was used to sample the Brillouin zone (BZ). The band structures were calculated using 50 k-points. In all calculations and for all species, the default *light* basis sets were used. All finite systems were sampled using 1 k-point (Gamma-point). Theoretical d$I$/d$V$ maps were calculated by the FHI-AIMS program package and with Probe Particle Scanning Probe Microscopy (PPSTM) code for an *s*-like orbital tip.

**Acknowlagement**

J. Lu acknowledges the support from MOE Tier 2 (MOE2019-T2-2-044) and MOE (Singapore) through the Research Centre of Excellence program (grant EDUN C-33-18-279-

V12,I-FIM). We acknowledge support from the Praemium Academie of the Academy of Science of the Czech Republic and the CzechNanoLab Research Infrastructure supported by MEYS CR (LM2018110). P.J., A.P., O.S. and S. E. acknowledge the support of the GACR 20-13692X. C. Chi acknowledges the support from MOE Tier 1 grant (R-143-000-B62-114) and Tier 2 grant (MOET2EP10120-0006).

**Author Contributions**

S.S. and J.L. conceived and designed the experiments. S.S., X.P., and J.S. performed the on-surface synthesis and LT-STM/AFM measurements. A.P., O.S. and P.J. performed the IETS measurements with NiCp2-tip. F.J.G. and A.L. assisted in the qPlus AFM measurements. P.W.N., J.W., and C.C. performed the organic synthesis of the precursors. S.E., Z.S., J.K. and P.J. performed the theoretical calculations. H.S., and C.S. helped to performed the LT-STM measurements. S.S., P.J. and J.L. wrote the manuscript with input from all authors. All authors contributed to the scientific discussion.

**Competing interests**

The authors declare no competing interest.

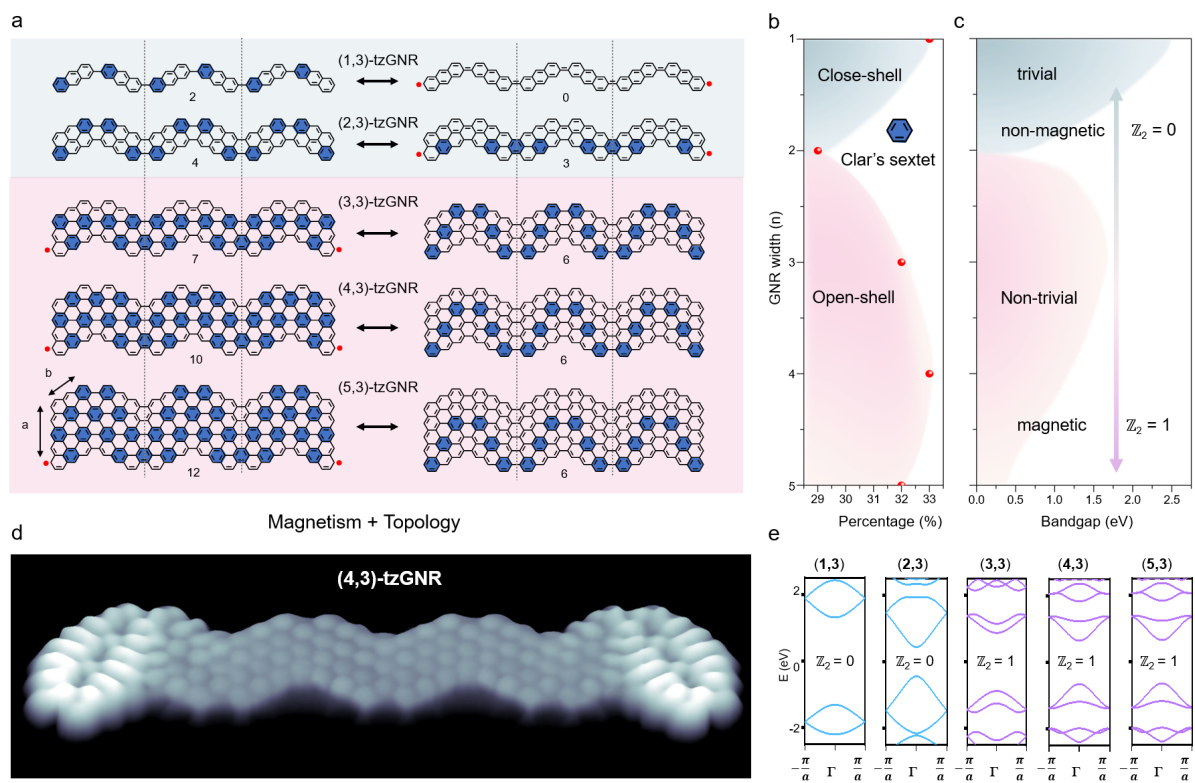

**Fig. 1 | Design concept and synthetic strategy of MT-GNRs. a**. Atomic structures of a series of (a,b)-tzGNRs (a and b denote the number of rings at each parallelogram nanographene that formed by the precursors). The dashed lines highlight the unit cells of the GNRs. The number of Clar's sextet in each unit cell of the closed-shell and open-shell tzGNRs is indicated below each corresponding resonance structure. **b.** The percentage of Clar's sextet as a function of the GNR width in each unit cell. **c.** Calculated band gap of the GNRs as a function of ribbon width which reveals a band gap reopening between (2,3)-tzGNR to (3,3)-tzGNR accompanied with a topological phase transition from trivial to non-trivial topological classes. **d**. A representative constant-current BR-STM image of an individual (4,3)-tzGNR with a length of eight monomer units ($V_s$ = 2 mV, $I_t$ = 200 pA). **e.** Calculated band structures of the ztGNRs with $\mathbb{Z}_2$ invariant determined from Zak phase calculation.

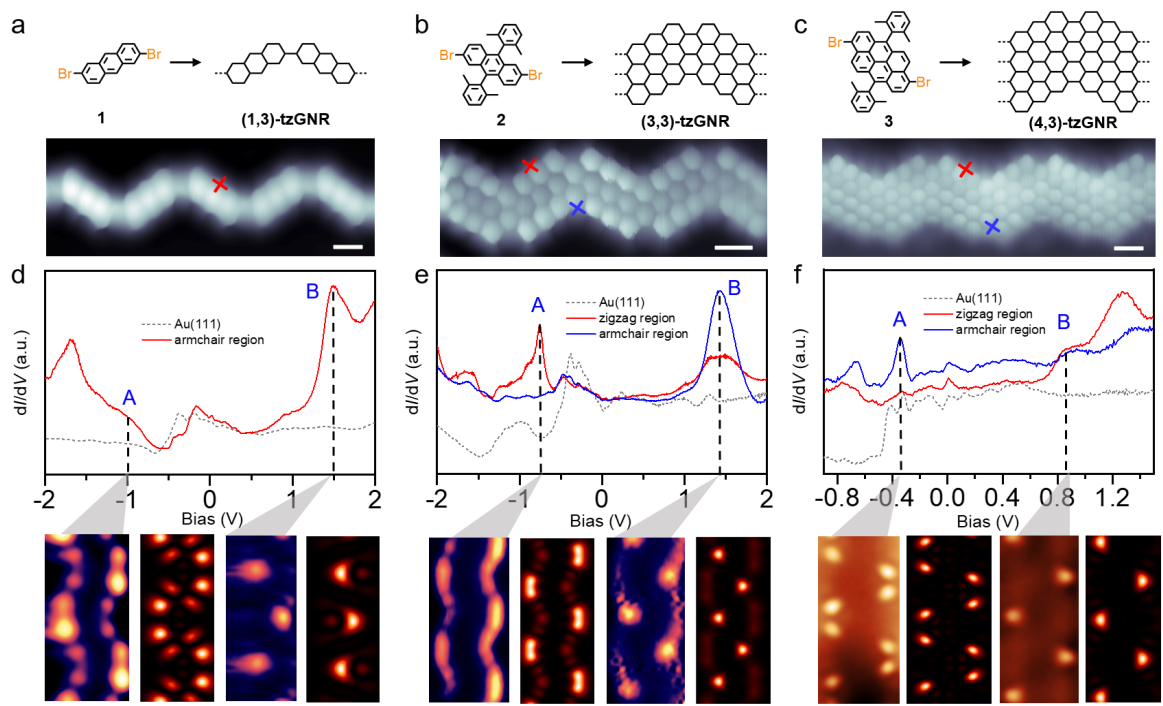

**Fig. 2 | Experimental and calculated electronic structure of the ztGNRs. a-c**. Synthetic route of (1,3)-, (3,3)-, and (4,3)-tzGNR with the corresponding constant height BR-STM images using a CO-functionalized tip ($V_s$ = 20 mV). All scale bars are 0.5 nm. **d-f**. Upper panel: d$I$/d$V$ point spectroscopy collected on edges of the GNRs (tip position marked by cross). The dashed black curve refers to the reference spectrum collected on bare Au(111). Lower panel: Constant current d$I$/d$V$ maps acquired at the energies corresponding to peaks A (VB) and B (CB). Simulated d$I$/d$V$ maps acquired at different energy positions corresponding to experimental ones are shown next to the experimental d$I$/d$V$ maps Note the lock-in parameters for d$I$/d$V$ and d$I$/d$V$ maps: modulation voltage $V_{ac}$ = 20 mV, $f$ = 479 Hz.

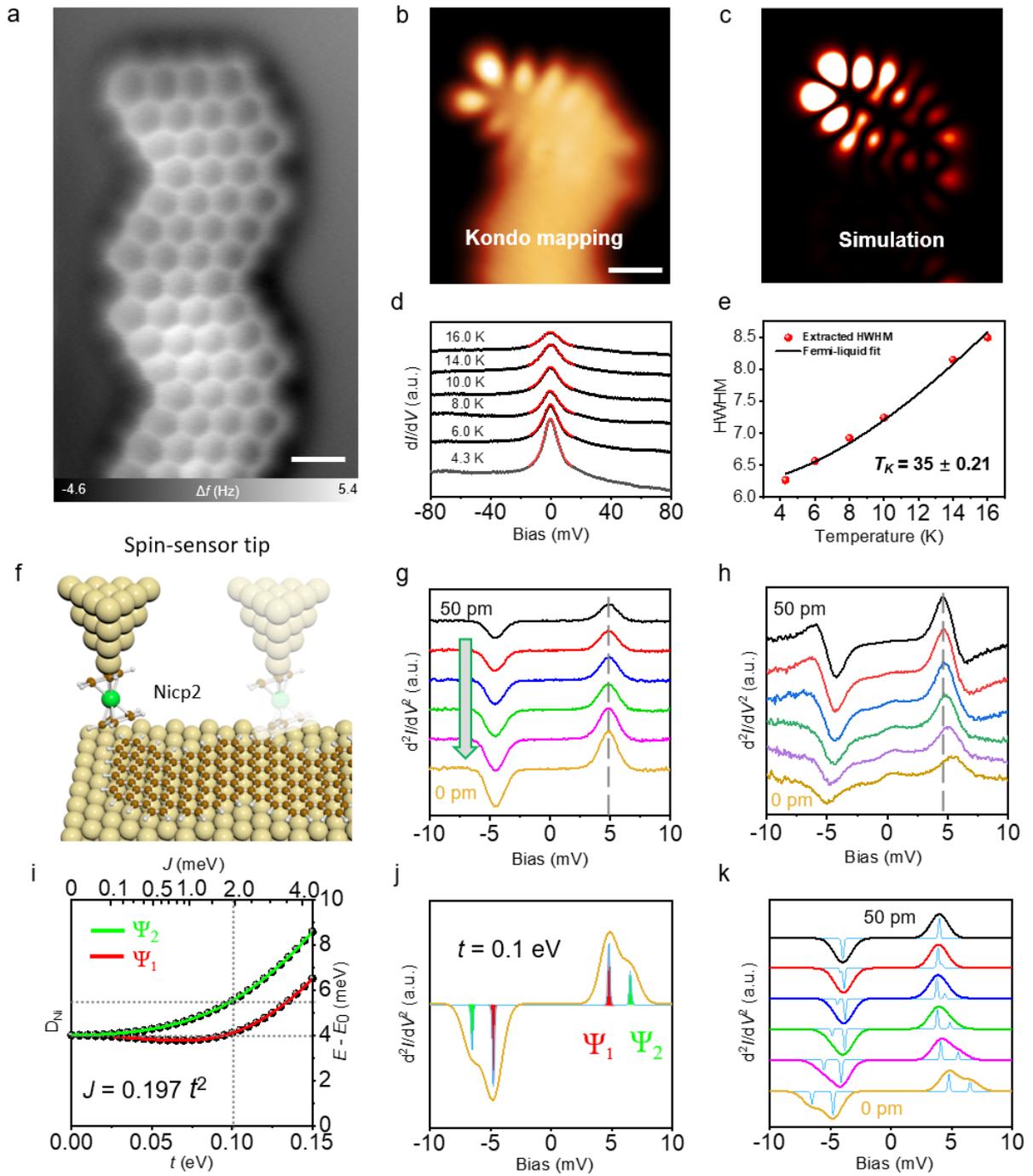

**Fig. 3 | Structure and magnetism characterization of the (4,3)-tzGNR. a**. ncAFM image of (4,3)-tzGNR terminal acquired with a qPlus sensor with a CO-functionalized tip ($V_S$ = 1 mV). **b**. Constant-height Kondo mapping of the GNR terminal ($V_s$ = 0 V). Scale bars for **a** and **b** is 0.5 nm. **c**. Simulated Kondo mapping of the GNR terminal (**b**). **d**. Temperature dependent d$I$/d$V$ spectrum taken over the terminal of the GNR ($V_s$ = 80 mV, $V_{ac}$ = 2 mV, and

*f* = 439 Hz). **e**. Extracted half-width at half-maximum of the Kondo resonance as a function of temperature (red points) together with Fermi-liquid model fitting (blue curves). **f**. Schematic illustration of probing the GNR magnetism with a nickelocene-functionalized tip. d$^2$I/dV$^2$ spectra acquired above the bulk (**g**) and the terminal (**h**) of the (4,3)-tzGNR at a distance between z = 0 and z =50 pm (The tip-sample distances are defined by the frequency shift of the Kolibri sensor, as shown in Fig. S5). **i**. Eigenenergies of the Hubbard model (black dots) and of the Heisenberg model (color lines) **j**. Calculated inelastic tunnelling spectrum at hopping parameter *t* = 0.1 eV with assigned origin of the individual peaks (respectively at Heisenberg exchange *J=2 meV*). $\Psi_1$ and $\Psi_2$ are defined in Extended Data Table 1. **k**. Caluclated inelastic tunnelling spectra at different hopping values, which show good agreement with the experimental spectra (h).

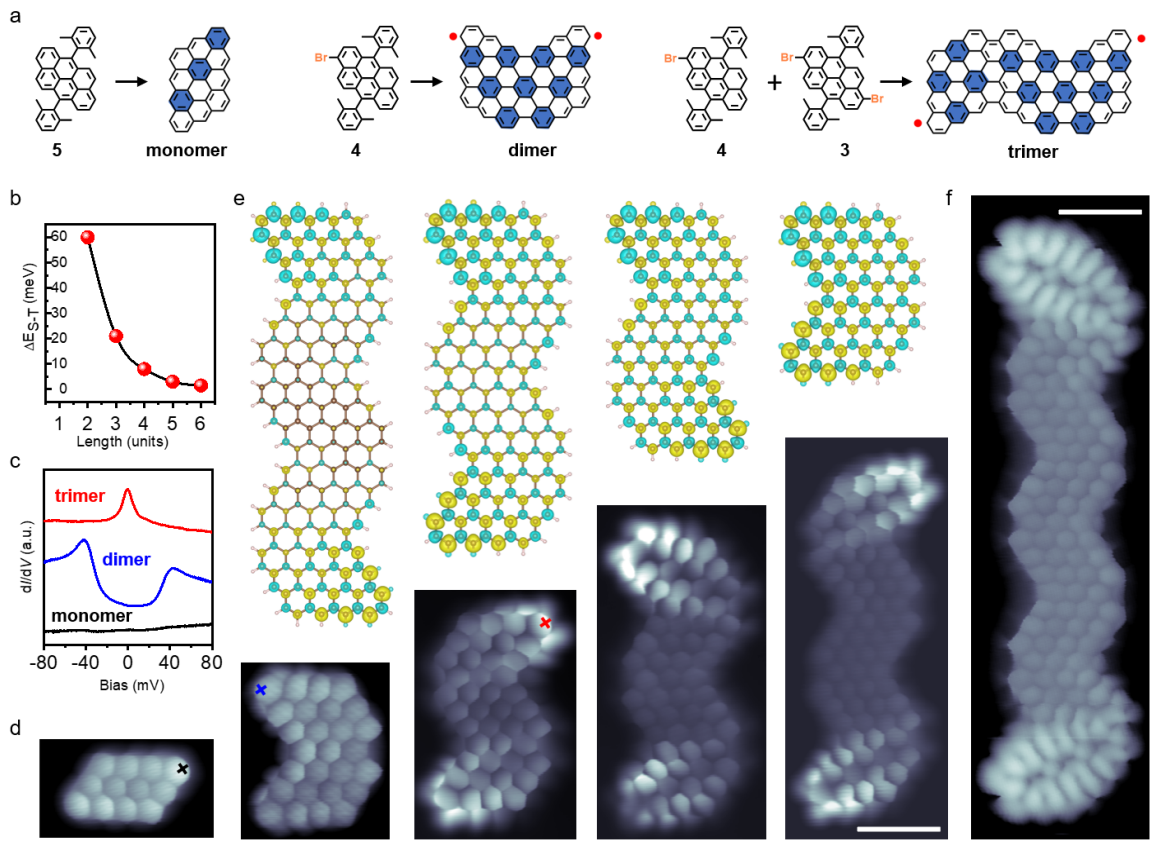

**Fig. 4 | Length-dependent terminal magnetic coupling of GNRs. a**. Synthetic scheme of the (4,3)-tzGNR monomer, dimer, and trimer from precursor **5**, **4**, and **4**+**3**, respectively. **b**. DFT calculated spin-flip excitation energies of free standing (4,3)-tzGNRs as a function of length. **c**. d$I$/d$V$ spectra taken over the monomer, dimer, and trimer where the positions are indicated by a black, blue, and red crosses over the BR-STM images in **d**. **d**. Constant height BR-STM images of the (4,3)-tzGNRs from monomer to pentamer ($V_s$ = 20 mV). Scale bar is 1 nm. **e**. DFT calculated spin density distributions of the (4,3)-tzGNRs from dimer to pentamer, which indicate the spins are delocalised over the terminal unit of the GNRs. **f**. Constant current BR-STM image of a (4,3)-tzGNR ($V_s$ = 2 mV $I_t$ = 200 pA), Scale bar is 1 nm. All the d$I$/d$V$ spectroscopy parameters are ($V_s$ = 80 mV, $V_{ac}$ = 2 mV, and $f$ = 439 Hz).

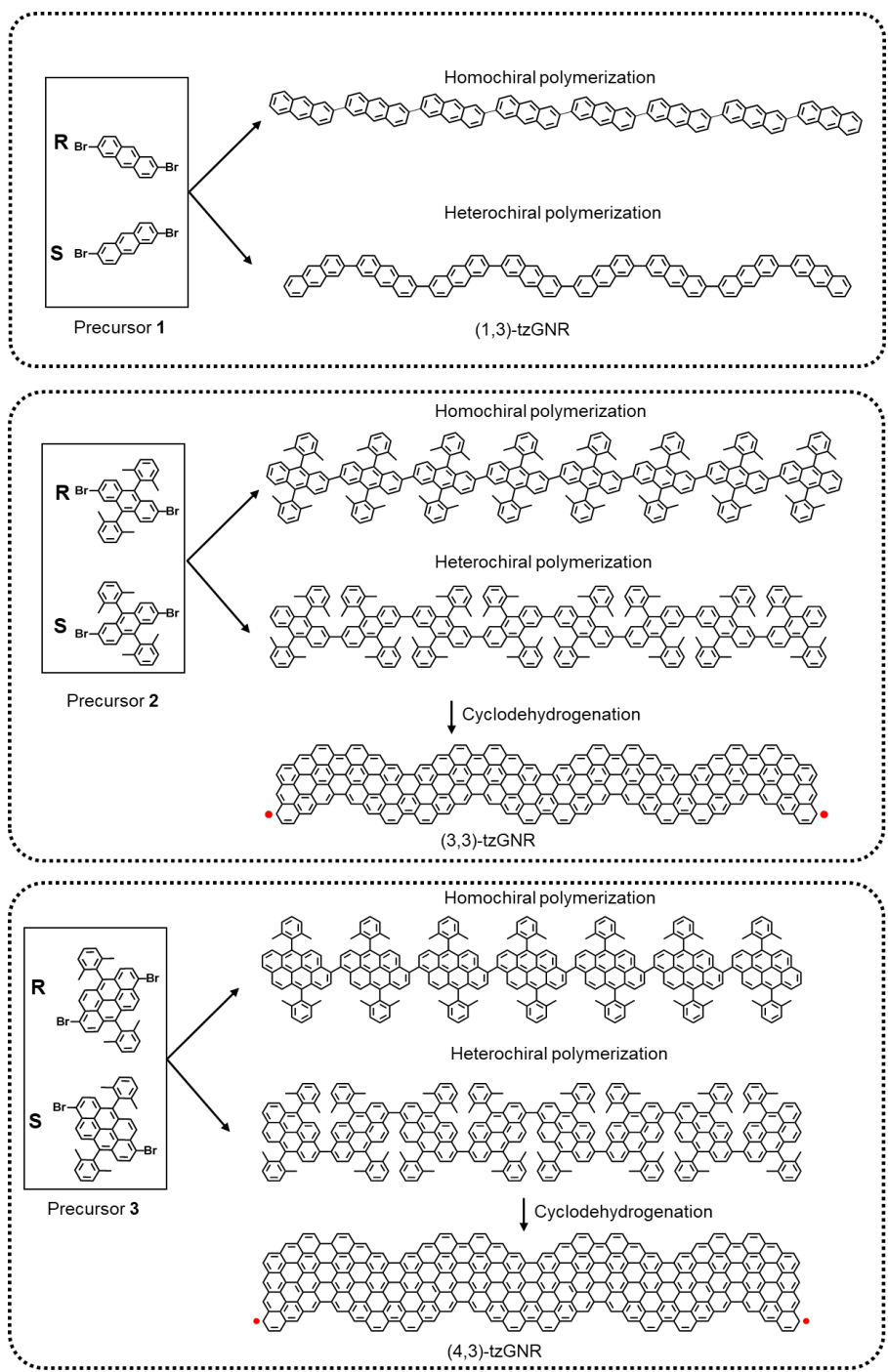

**Extended Data Fig. 1 | On-surface synthetic scheme of the GNRs.** Two possible polymerization approaches (namely, homochiral polymerization and heterochiral polymerization) from the two enantiomers of the precursors. Only the heterochiral polymerization and subsequent cyclodehydrogenation can lead to the target GNRs.

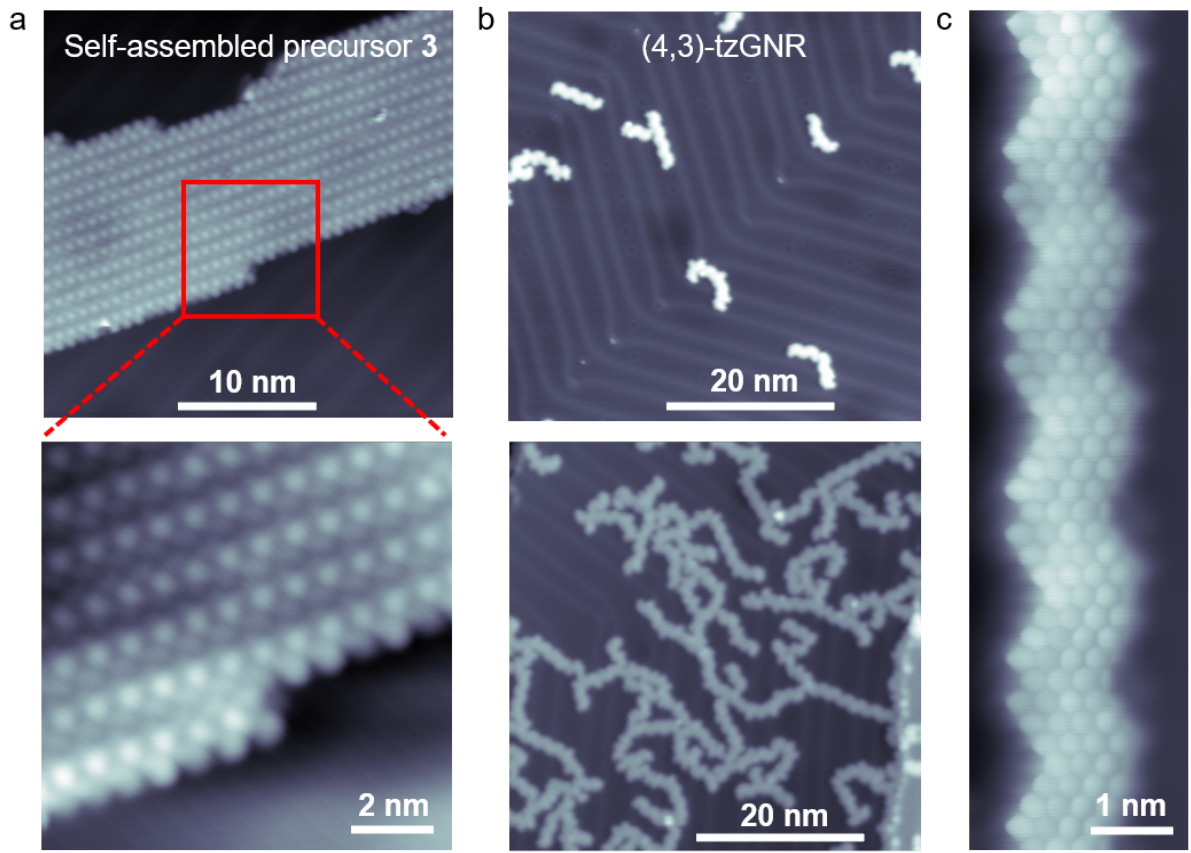

**Extended Data Fig. 2 | On-surface synthesis of the (4,3)-tzGNR. a**, Self-assembled precursor **3** in large scale (upper panel) and zoomed in island (lower panel). **b**, GNRs in different coverage ratios, where individual shorter GNRs were obtained at smaller coverage (upper panel), connected longer GNRs were obtained at larger coverage (lower panel). **c**. BR-STM image of an (4,3)-tzGNR bulk with 10 monomer unit length.

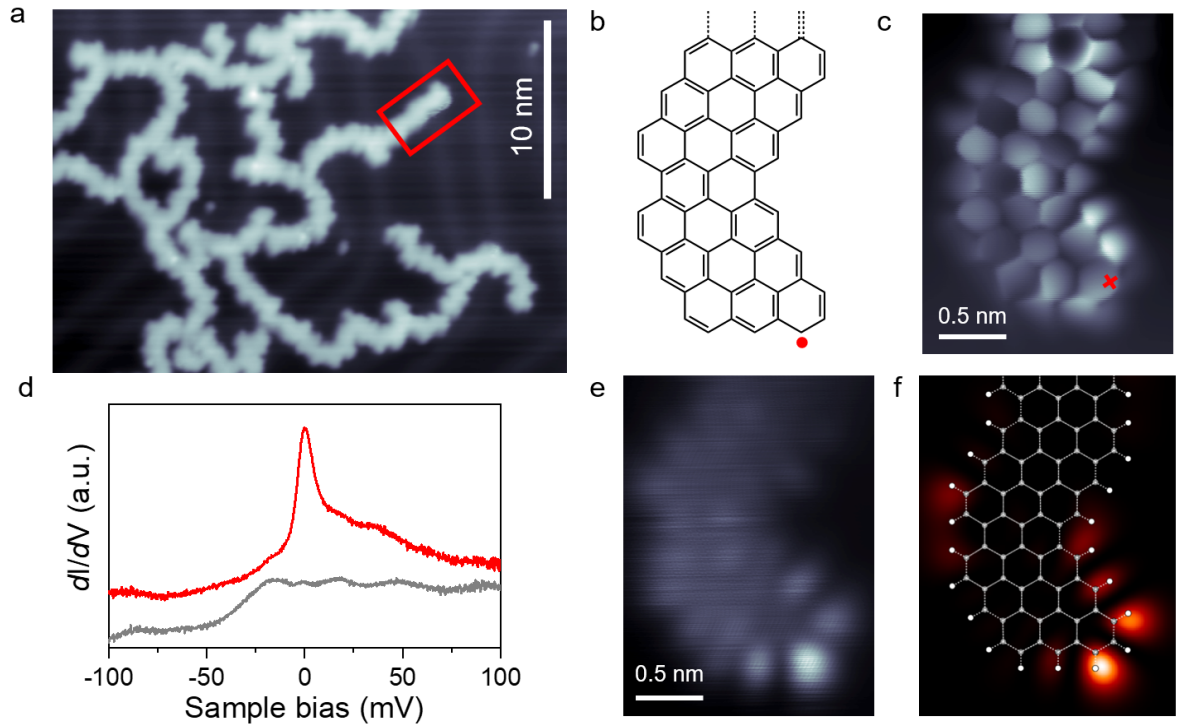

**Extended Data Fig. 3 | On-surface synthesis of (3,3)-tzGNR. a**, Large scale STM image of the synthesized 3GNR on Au(111). **b**, Molecular structure of the 3GNR terminal which show an unpaired electron. **c**, BR-STM image of the (3,3)-tzGNR terminal. **d**, d$I$/d$V$ spectrum taken over the terminal of (3,3)-tzGNR (marked as red cross in **c**) which shows a Kondo resonance peak. ($V_s$ = 80 mV, $V_{ac}$ = 2 mV, and $f$ = 439 Hz) **e**, Kondo mapping of the (3,3)-tzGNR terminal. **f**, Simulated Kondo mapping of the 3GNR terminal with molecular model overlay.

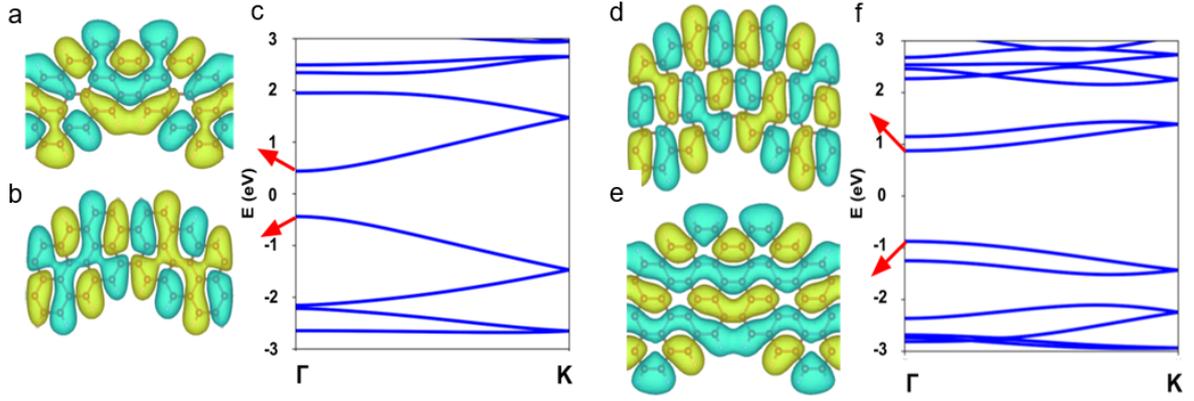

**Extended Data Fig. 4 |** Band structure plots of (2,3)-tzGNR **c** and (3,3)-tzGNR **f**, accompanied by plots of the wave functions of the frontier bands at the Gamma-point. **a**, (2,3)-tzGNR' conduction band, **b**, (2,3)-tzGNR' valence band, **d**, (3,3)-tzGNR' conduction band, and **e**, (3,3)-tzGNR' valence band. Note the change in the characters of the CB and VB wave functions from bonding (antibonding) to antibonding (bonding) in the CB (VB) represents a topological phase transition from (2,3)-tzGNR to (3,3)-tzGNR.

**Extended Data Table 1 |** Characteristics of the eigenstates of the Heisenberg model for the $d^2I/dV^2$ simulation NiCp2-tip. Note that only $m_{tot}$ is a conserved quantum number. The state $\Psi_0$, $\Psi_1$ and $\Psi_2$ represent the ground state, the spin flip at the Ni site, respectively.

| state | $m_{Ni}$ | $m_{obj}$ | $m_{tot}$ | E |
|---|---|---|---|---|
| $\Psi_2$ | $\pm 1$ | $\pm \frac{1}{2}$ | $\pm \frac{3}{2}$ | $D + \frac{1}{2}J$ |
| $\Psi_1$ | $\pm 1$ | $\pm \frac{1}{2}$ | $\pm \frac{1}{2}$ | $\frac{1}{4}\left(2D - J + \sqrt{((2D-J)^2 + 8J^2)}\right) \approx D - \frac{J}{2} + \frac{J^2}{2D}$ |
| $\Psi_0$ | 0 | $\pm \frac{1}{2}$ | $\pm \frac{1}{2}$ | $\frac{1}{4}\left(2D - J - \sqrt{((2D-J)^2 + 8J^2)}\right) \approx -\frac{J^2}{2D}$ |

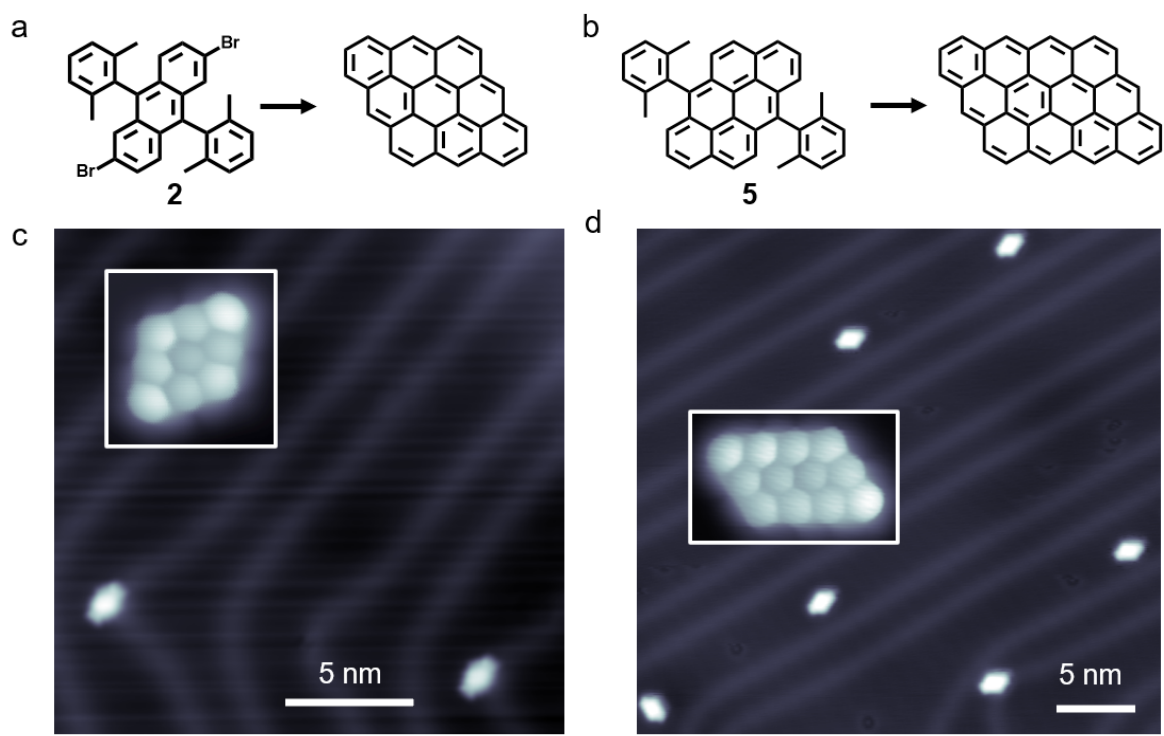

**Extended Data Fig. 5 | On-surface synthesis and characterization of the (3,3)-tzGNR and (4,3)-tzGNR monomers. a,b**, On-surface synthesis of (3,3)-tzGNR and (4,3)-tzGNR monomers from precursor **2** and **5,** respectively. **c,** STM image of the (3,3)-tzGNR monomer. ($V_s$ = -1 V, $I_t$ = 50 pA). Inset is the BR-STM of the (3,3)-tzGNR monomer. **d**, STM image of (4,3)-tzGNR monomers. ($V_s$ = 1.5 V, $I_t$ = 100 pA). Inset is the BR-STM of (4,3)-tzGNR monomer.

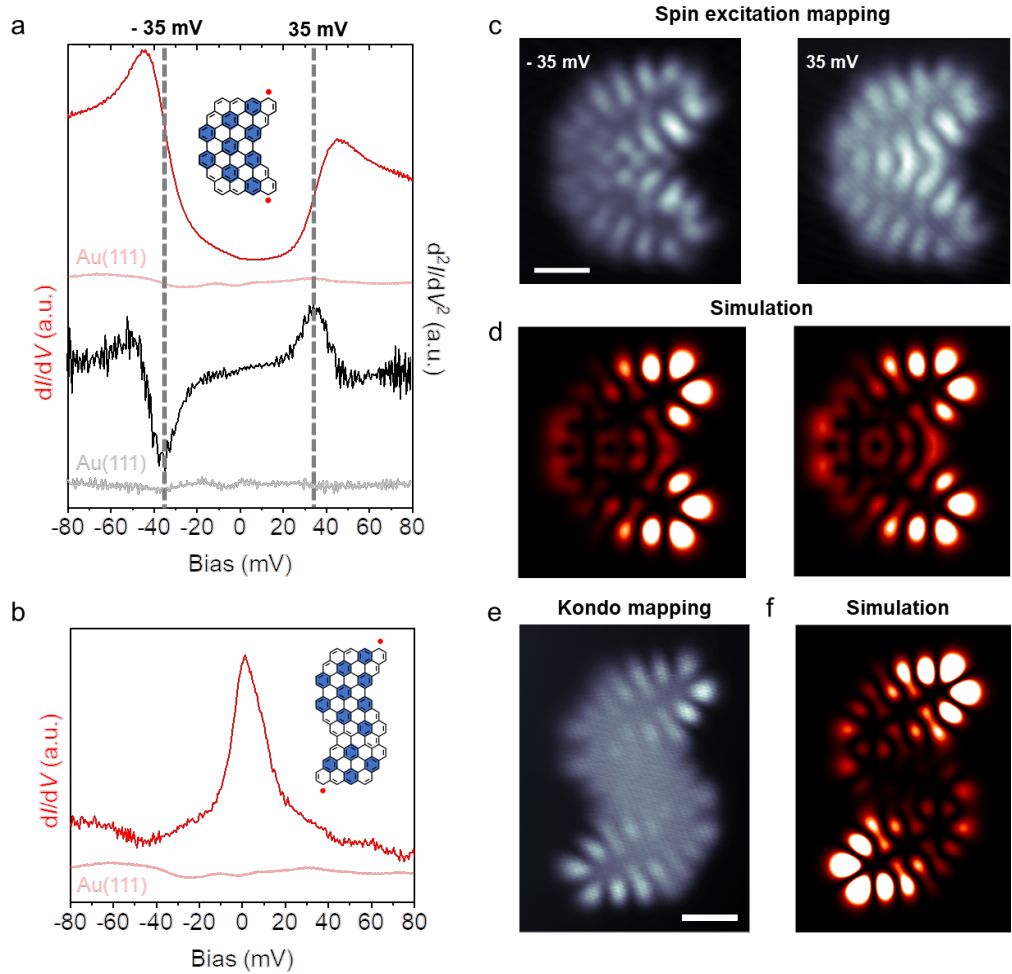

**Extended Data Fig. 6 | Electronic and magnetic characterization of the (4,3)-tzGNR dimer and trimer. a**, d$I$/d$V$ and d$^2I$/d$V^2$ spectra of the dimer. **b**, d$I$/d$V$ spectrum of the trimer. ($I_t$ = 1.5 nA, $V_{ac}$ = 2 mV, and $f$ = 439 Hz). **c**, d$I$/d$V$ maps acquired at the energy of ± 35 mV. ($I_t$ = 1.5 nA, $V_{ac}$ = 2 mV). **d**, Simulated d$I$/d$V$ maps at the energies corresponding to **a**. **e**, Experimental and **f**, simulated d$I$/d$V$ maps of the trimer at the energy of 0 mV. ($I_t$ = 1.5 nA, $V_{ac}$ = 2 mV, and $f$ = 439 Hz). All scale bars are 0.5 nm.

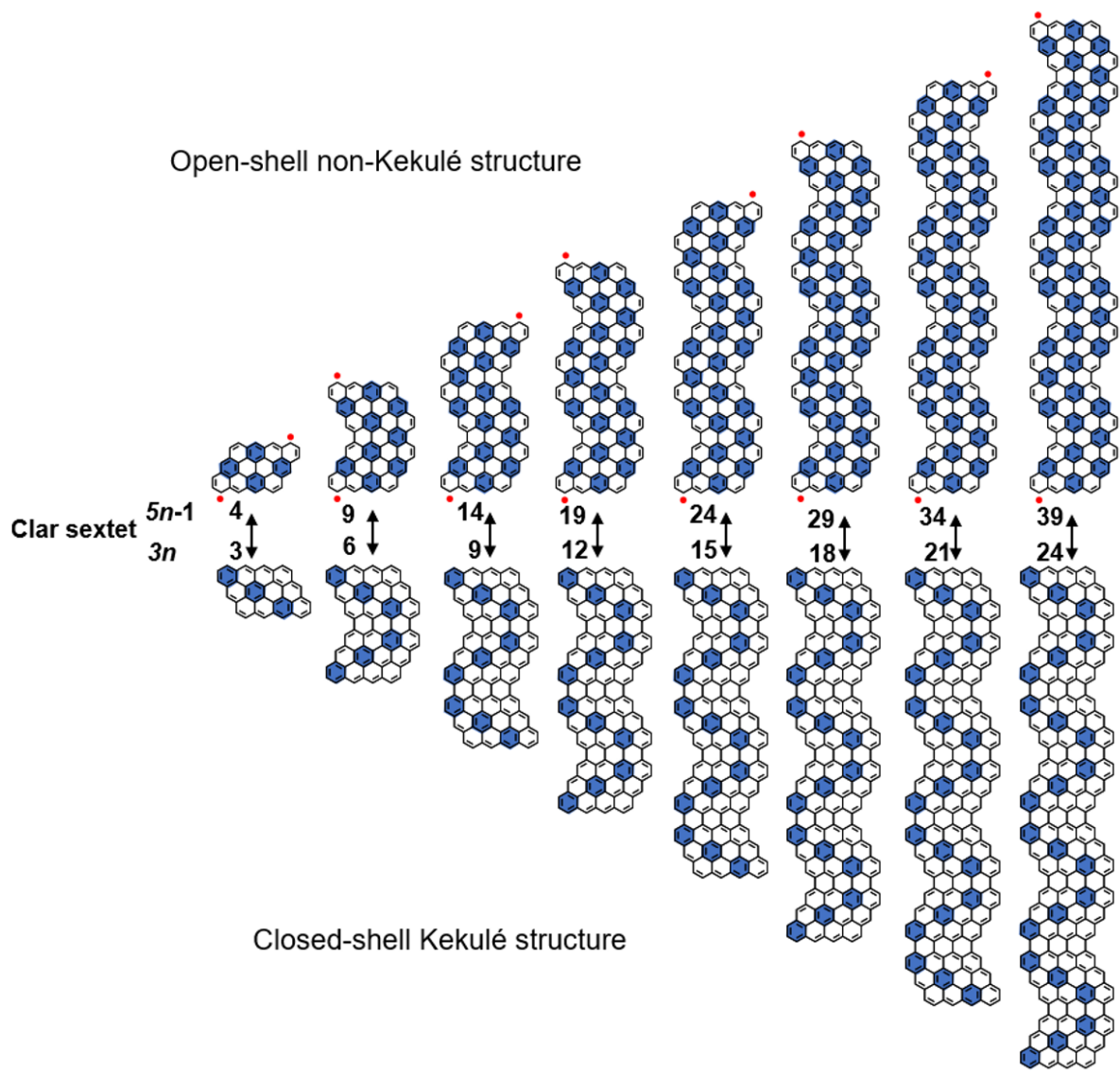

**Extended Data Fig. 7 | Open-shell Kekulé structure and closed-shell Kekulé structure of the (4,3)-tzGNR with finite lengths.** The red dot represents radical site in the open-shell Kekulé structure of the GNRs. Blue filled rings correspond to the Clar's π sextets. The closed-shell Kekulé structure of the GNRs can be represented with 3n of migrating aromatic π sextet (n is the number of monomer unit); while the corresponding open-shell non-Kekulé structure possesses a (5n – 1) of Clar's π sextets. The extra Clar's π sextet obtained favors the formation of unpaired electrons at the terminals.